\documentclass{article}
\usepackage{graphicx}
\usepackage[affil-it]{authblk}

\begin{document}

\title{Astrometric detection feasibility of \\ 
gravitational effects of quantum vacuum }

\author{M. Gai
\thanks{Electronic address: \texttt{gai@oato.inaf.it}; Corresponding author}, 
A. Vecchiato}

\affil{ Istituto Nazionale di Astrofisica - Osserv. Astrofisico 
di Torino, \\ 
V. Osservatorio 20, 10025 Pino T.se (TO), Italy }



\maketitle
\begin{abstract}
This work analyzes in some detail the feasibility of testing with
astrometric measurements the hypothesis that Quantum Vacuum can have
gravitational effects, as suggested in a series of recent papers 
(\cite{2012Ap&SS.339....1H,hajdukovic:hal-00905914,hajdukovic:hal-00965289}).
In particular, the possibility of detecting an excess shift of the
longitude of the pericenter in the orbit of the trans-neptunian system
UX25 and its satellite is investigated. The excess shift which might
be experimented by the orbit of the satellite was estimated, under
reasonable working hypothesis, to be about $0.23$~arcsec per orbit.
Several observing scenarios are explored here, including those using
conventional and adaptive optics telescopes from ground, and some
spaceborne telescopes. 
The preliminary assessment of astronomical detection of quantum vacuum 
provides encouraging results: the best currently available instrumentation 
seems to be able, over a time frame of a few years (due to the need 
of cumulating the precession effects), to reach unambiguous results. 

\end{abstract}

\section*{Introduction}

In a recent paper Hajdukovic \cite{hajdukovic:hal-00905914} speculated
about the possibility that Quantum Vacuum has to be included in a
complete and correct description of Gravity. Such ``speculation''
stems from the simple consideration that: a) the presently accepted
model of the gravitational interaction can agree with the current
experimental findings only by assuming the existence of entities like
Dark Matter (DM) and Dark Energy (DE); b) there do exist another entity,
namely the Quantum Vacuum (QV) whose existence is essential in the
context of Quantum Physics for the Standard Model of Particle Physics
to be in agreement with the experiments; c) the nature and the properties
of DM and DE are currently unknown, and all the experimental attempts
which tried to find evidences of the existence of any of their candidates
have failed up to now. Given these premises, it is not unreasonable
to assume that QV, DM, and DE can actually be the same entity or,
in other words, that QV has the same gravitational effects that now
are attributed to DM and DE.

This is actually not a completely new attempt. The existence of DE
was postulated after the measurement of an accelerated expansion of
the Universe at high redshifts, however, from a qualitative point
of view, such kind of expansion can also be reproduced by a non-zero
cosmological constant $\Lambda$, interpreted as the effect of a homogeneous
fluid with constant negative pressure. In principle, it would be natural
to identify such fluid with the QV because there is a mechanism in
Quantum Field Theory (QFT) to produce a negative pressure. This elegant
idea didn't work because of the so-called \emph{cosmological constant
problem} \cite{1989RvMP...61....1W} i.e. because the energy of the
QV predicted by the QFT was many orders of magnitude larger than the
one which could be attributed to $\Lambda$ from astrophysical measurements.

Stemming from the same premises, recently it was suggested that, on
the assumption that particles and antiparticles forming the QV have
opposite gravitational charge, the QV energy could be compatible with
that attributed to DE \cite{2012Ap&SS.339....1H} and that the same
hypothesis could be used to explain the effects now attributed to
DM. From the astronomical point of view, however, the consequences
of attributing a gravitational effect to QV should in principle be
considered at all scales, not just at the cosmological ones needed
to explain the DE effect. In the simple context of the two body problem,
e.g., the additional gravity source represented by QV would necessarily
induce a deviation from the perfect symmetry of the gravity field
which is necessary to have closed orbits, and therefore show itself
up as an excess of the shift of the pericenter of a body orbiting
around a central gravity source.

In the Solar System such deviations can be already induced by the
Sun itself in a classical context, but they are negligible far from
it, i.e. at the typical distances of the trans-neptunian objects,
while those attributed to the QV energy would not decrease. This explains
the potential relevance of accurate astrometric monitoring of trans-neptunian
objects claimed in \cite{hajdukovic:hal-00908554} as a target for
Fundamental Physics tests, in particular for the detection of possible
non-Newtonian contributions to their orbits.

Here we are exploring in more detail the feasibility of such kind
of measurements for the UX25 trans-neptunian binary system.

\section{The target system: 2002 UX25}

The trans-Neptunian object (55637) 2002 UX25 (hereafter, UX25 for
short) orbits the Sun in the Kuiper belt (semi-major axis 42.869 AU)
beyond Neptune, with period 280.69 years, and it has a single known
moon with fairly large size. A more detailed description of its characteristics
is available in the literature \cite{2013ApJ...778L..34B}. UX25 is
located at about 43 AU from the Sun, and has visual magnitude close
to 20~mag. The primary and secondary diameters are estimated to be
about $670\, km$ and $190\, km$ (from spherical model of thermal
emission). The angular diameters are then respectively $21.4\, mas$
(primary) and $6.1\, mas$ (secondary), mostly unresolved even by
AO telescopes. The magnitude difference between primary and secondary
is 2.7 mag (a factor 12 in brightness), fairly consistent with the
area ratio $\sim12.4$ associated to the estimated diameter ratio
$\sim3.32$, assuming similar composition and surface structure. The
mass ratio is 43.85, and the semi-major axis of orbit
is 4'770 km, corresponding to about $152\, mas$. 

The binary period is 8.3~days, and the eccentricity is $\epsilon=0.17$.
The orbit inclination is close to $60^{\circ}$, so that it is closer
to the edge-on case than to the face-on case, but still well resolved
in two dimensions. 

As a binary system located in a remote region of the solar system,
it is affected by small perturbations from other bodies, making it
a convenient candidate for the quantum vacuum detection experiment.
The drawback of the source is that its magnitude is not very bright,
being slightly fainter than 20 mag for the primary, and close to 23
mag for the secondary. 

\begin{figure}
\begin{centering}
\includegraphics[scale=0.47]{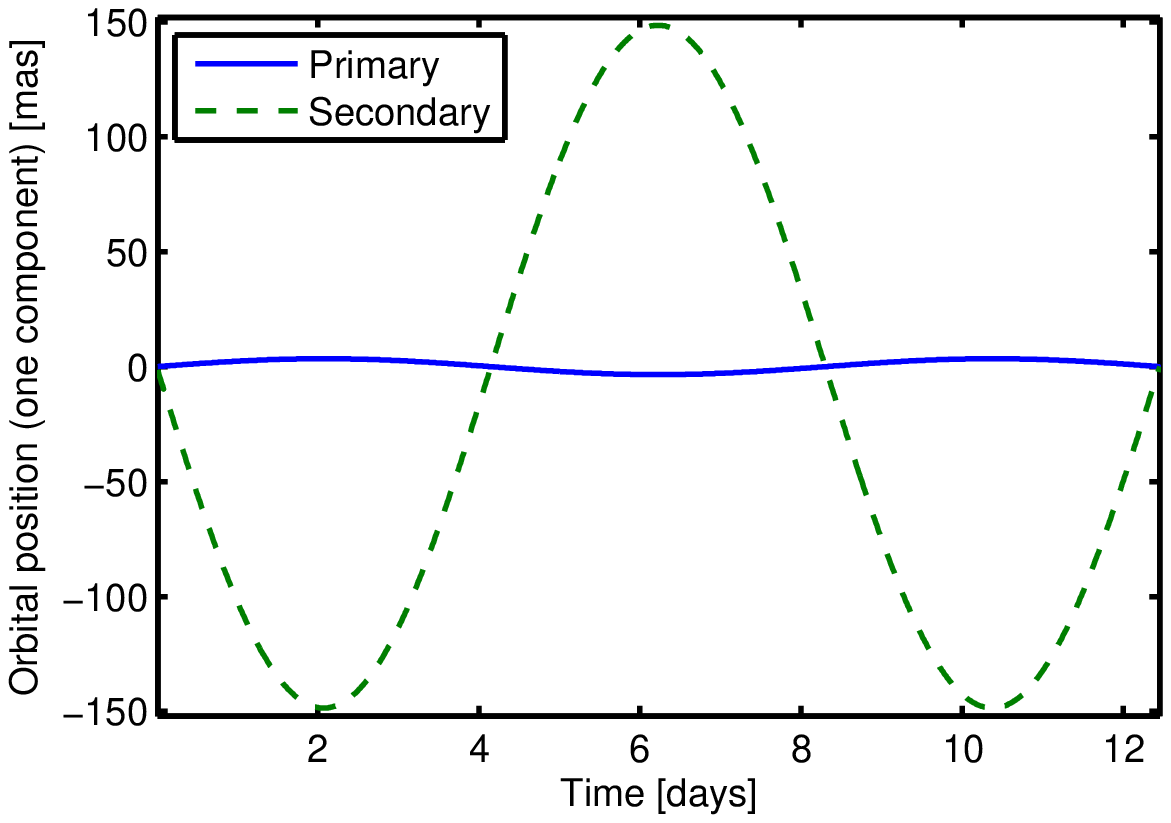}
\includegraphics[scale=0.40]{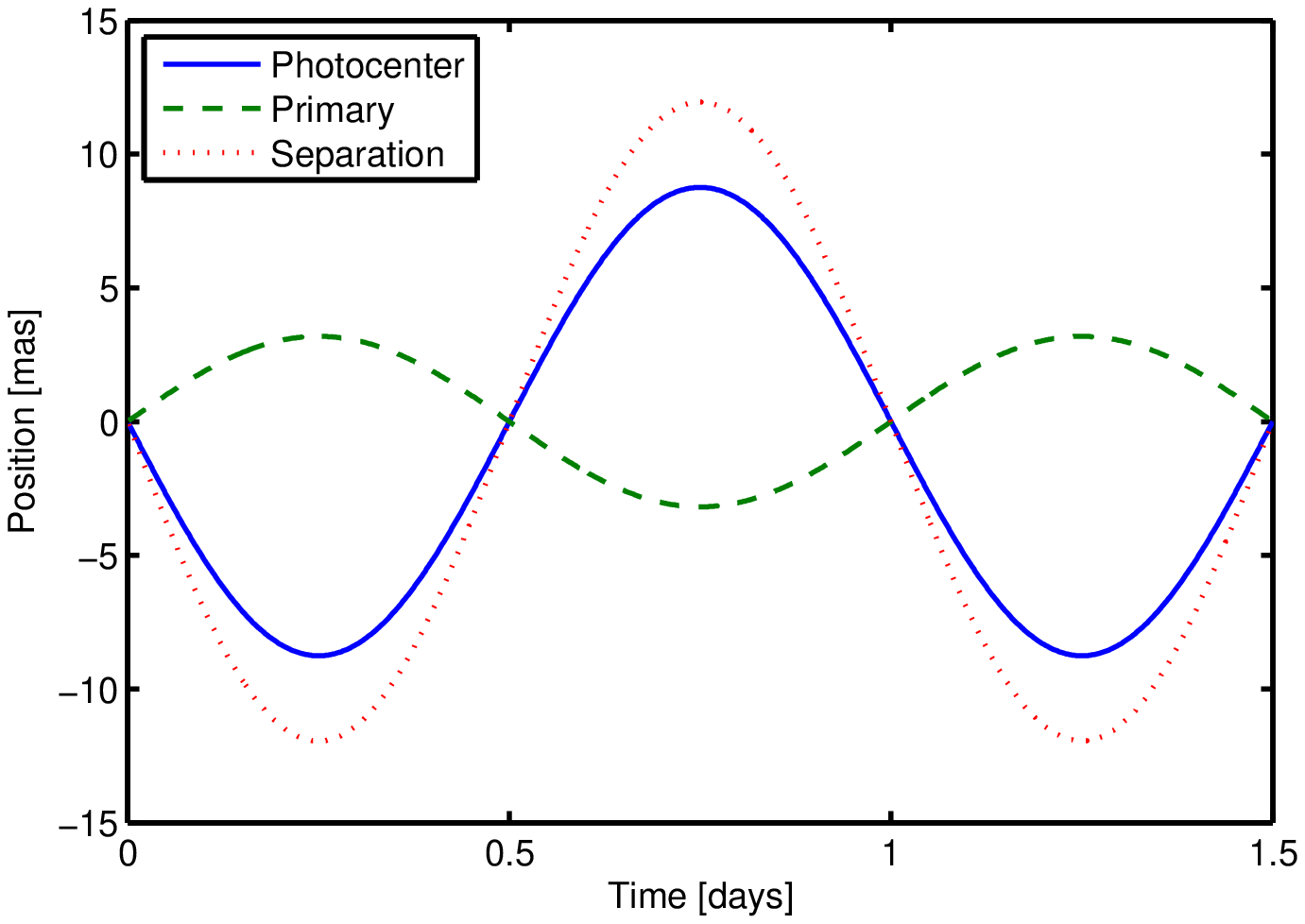}
\par\end{centering}
\caption{\label{fig:ObjPos_Phot}One-dimensional object position (left) and
photocenter position (right) over 1.5 orbital periods}
\end{figure}

In the approximation of circular orbits, the relative one-dimensional
position of the two objects over a time elapse of 1.5 periods is shown
in Fig.~\ref{fig:ObjPos_Phot} (left). The photocenter motion for
each component, and the overall photocenter, is shown in the right
panel. This assumes that the two objects are unresolved, i.e. observed
as a single spot of light, e.g. by a conventional ground based telescope. 

The plot on the left panel in Fig.~\ref{fig:ObjPos_Phot} is representative
of what could be achieved by an AO-equipped telescope, able to resolve
in most observations the primary and secondary components of the binary
object UX25.

\subsection{Astrometric signal amplitude }

\label{sub:AstromSigAmpli}
\begin{figure}
\begin{centering}
\includegraphics[scale=0.7]{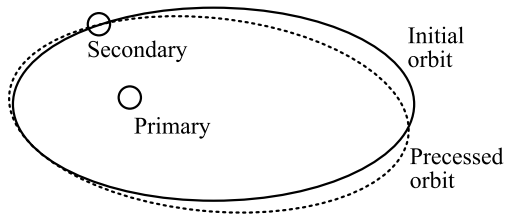}
\par\end{centering}

\caption{\label{fig:Orbit1}Representation of orbit precession (face-on view,
not to scale)}
\end{figure}

\begin{figure}
\begin{centering}
\includegraphics[scale=0.5]{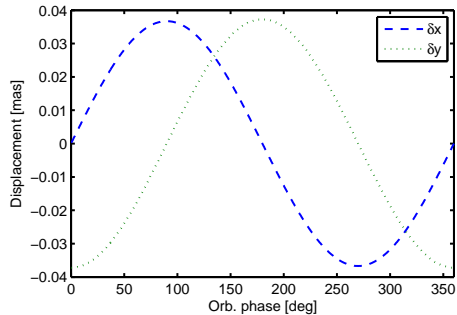}
\par\end{centering}

\caption{\label{fig:AstrSignal1}Amplitude of the astrometric signal associated
to five year precession of the 2002 UX25 orbit}
\end{figure}

A qualitative representation of the orbit
recession phenomenon is shown in Fig.~\ref{fig:Orbit1}. The position
of the periastron, and of any other point in the orbit at a given
time, evolve continuously with time, so that the orbit is no longer
ckosed as in the standard Keplerian solution of the two-body problem.
The amount of precession accumulated over five years, corresponding
to about 220 orbits, is $50".55$. This corresponds to a peak position
displacement on the sky of a few ten micro-arcsec ($\mu as$), as
shown in Fig.~\ref{fig:AstrSignal1}. \textbf{In order to achieve
a clear detection at the $3\sigma$ level, then, we need a measurement
precision equivalent to about $15\,\mu as$. }

The amplitude is rather small, however all orbital positions are displaced
in a correlated way, so that the goal appears to be compatible with
state of the art astrometry from the ground {[}ref. Cameron 2009{]},
in particular with the use of large telescopes equipped with adaptive
optics (AO). 

Given the high competition for time allocation of large telescopes,
it is necessary to formulate a good observing proposal, with strong
(astro-)physical motivations, and a robust implementation plan. Besides,
it is mandatory to demonstrate that the use of such facility is unavoidable,
i.e. that the experiment cannot be carried on with smaller instrumentation.
It is convenient to check the feasibility from ground based telescopes
of small class, limited by atmospheric turbulence, before addressing
large, adaptive optics (AO) equipped telescopes or space telescopes.

\subsection{Experimental approach }

\label{sub:ExpApproach}The first step is detection of a relevant
precession; the second is the determination of its amplitude, in order
to set proper bounds to the theory. In both cases, it is not convenient
to evaluate just a single orbital position (which would anyway be
impractical). The proposed approach consists in a set of observations
mapping the whole orbit, in order to evidence its variation over a
given period of time by comparing its basic geometric properties at
different epochs. 

It should be noted that the precision on our current knowledge of
the UX25 system is sufficient to mark it as a convenient candidate
for our study, but it is far from the level required by the experiment.
In practice, this means that the initial orbital state must be measured
with precision consistent with the experiment goal, as well as the
final state. Moreover, the orbit must be monitored also throughout
the intermediate period, although possibly with somewhat less demanding
requirements, in order to detect potential disturbing effects to be
modeled and removed. 

The initial orbit is suposed to be known with sufficient precision.
After a given time elapse, in which the periaster precession takes
place, a set of position measurements is performed to estimate the
new set of orbital parameters. 

The first conceivable test is that of consistency of the new orbit
with the initial one, i.e. the null test of no precession. Astrometric
noise on the primary and secondary photo-centres may be induced by
the real, non-spherical shape of the bodies, variations in the surface
structure and reflectivity (albedo), rotation and so on. Accurate
characterisation of the binary system is therefore required. 

The measurement can be based either on the separation of the two components
of the binary system, i.e. on the quantities shown in Fig.~1 (left),
or on the separation between primary and photocenter position, represented
in Fig.~1 (right). In the former case, the separation is roughly
one order of magnitude larger, but the measurement is mainly limited
by the SNR of the secondary. In the latter case, both primary and
photocenter have comparable precision, corresponding to the overall
SNR, but the separation is about one order of magnitude smaller. Moreover,
it can be shown from current simulations that the precision on the
secondary is about a factor three worse than that on the primary,
when the two components are resolved, but the degradation becomes
about one order of magnitude when the images are significantly superposed,
i.e. at least over a significant fraction of the orbital period. The
choice of data processing approach, from the same imaging observations,
is therefore not clear cut.

\section{Atmospheric limitations / characteristics}

Atmospheric turbulence is one of the main limitations of ground based
astrometry. However, in good observing conditions, and adopting thorough
control of systematic errors, it was possible to achieve precision
levels as good as $0.15\, mas$ \cite{1996ApJ...465..264P}, referred
to one hour exposures, using optimised conventional telescopes. Adaptive
optics may provide even better results, over very small fields, e.g.
order of $100\,\mu as$ over two minute exposures \cite{2009AJ....137...83C},
and optical long baseline interferometry is progressing towards the
$10\,\mu as$ goal (e.g. the ESO VLTI PRIMA facility, or the Keck
Interferometer). 

For the application currently investigated, it is convenient to extrapolate
the results from Pravdo and Shaklan (1996)\cite{1996ApJ...465..264P},
in particular concerning \textit{field averaging} of the atmospheric
noise. The one-dimensional error $\epsilon$ on the estimated position
of a target with respect to a set of $N$ reference stars in the field,
located at distances $d_{01}$ from the target along the measurement
direction, from an exposure of $T$ seconds, with a telescope having
diameter $D$, has variance 
\[
\epsilon^{2}=\frac{2.9\times10^{-5}}{N^{2}T}\left[\frac{1}{D^{4/3}}\sum_{m,n=1}^{N}\left(d_{0m}^{2}-\frac{1}{2}d_{mn}^{2}\right)-192\frac{1}{D^{2}}\sum_{m,n=1}^{N}\left(d_{0m}^{8/3}-\frac{1}{2}d_{mn}^{8/3}\right)\right]\,,
\]
where $d_{mn}$ are the separations among the reference stars, and
the result is in $arcsec^{2}$. The equation was derived from a simple
model \cite{1980A&A....89...41L}, assuming a Kolmogorov turbulence
spectrum with profile described according to Hufnagel (1974), and
a seeing parameter $r_{0}=10\, cm$. The parameters correspond to
a good quality observing site. All distances are in radians. 

It may be noted that the limiting error can be easily computed for
any field, in given observing conditions, providing results highly
dependent on the star separations. In practice, it may be convenient
to limit the range for locating reference stars around a given target,
since further objects at a longer distance may not help significantly
in improving on the error. Also, the residual atmospheric noise depends
on the telescope diameter $D$. 

\begin{figure}
\begin{centering}
\includegraphics[scale=0.55]{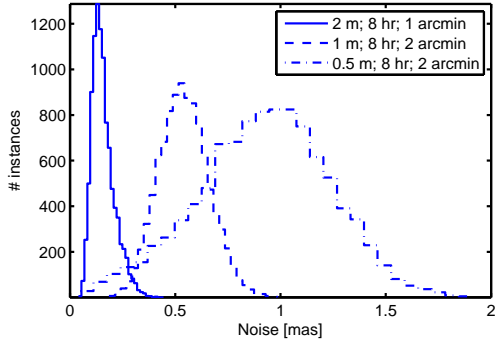}
\par\end{centering}
\caption{\label{fig:ResNoise_1night}Residual atmospheric noise over one night}
\end{figure}

The model is used to assess the potential performance of astrometric
observations from ground, by means of small telescopes (primary mirror
diameter $D=0.5\, m$ and $D=1\, m$), with exposure time of five
minutes and one hour, assuming the availability of $N=10$ stars within
a field of either $2'$ or $5'$. The target is assumed to be located
between the centre of the field and the mean of the reference star
photocentres. A set of $N_{I}=10,000$ random star positions, with
a uniform distribution, has been generated for each condition, with
results evidenced by the histograms in Fig.~\ref{fig:ResNoise_1night}.
Favourable conditions of higher stellar density may result in further
precision improvement. Moreover, the measurement can, in principle,
be averaged over a whole night (eight hours) of observations, providing
the results shown in Fig.~\ref{fig:ResNoise_1night}, including the
case of a $D=2\, m$ telescope with $N=10$ reference stars within
a field of $1'$.

\section{Assessment of orbit determination from conventional telescopes}

\begin{figure}
\begin{centering}
\includegraphics[scale=0.5]{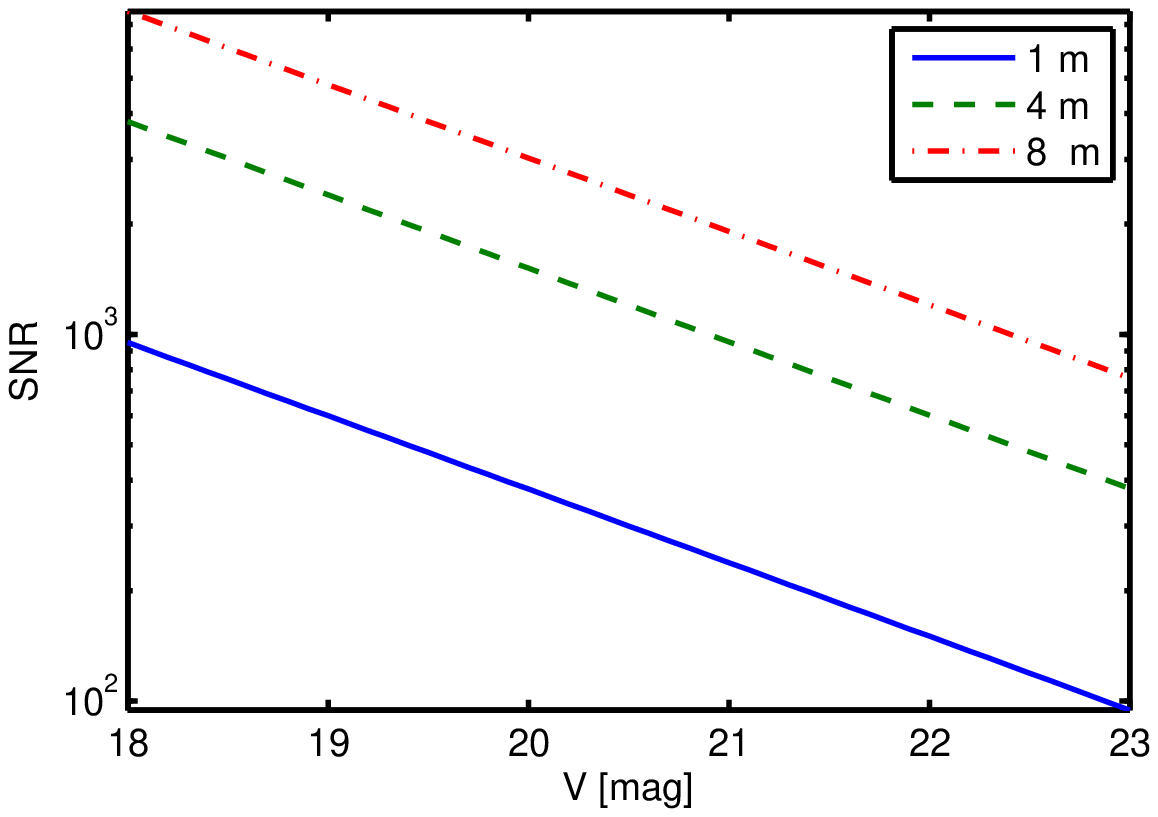}
\includegraphics[scale=0.5]{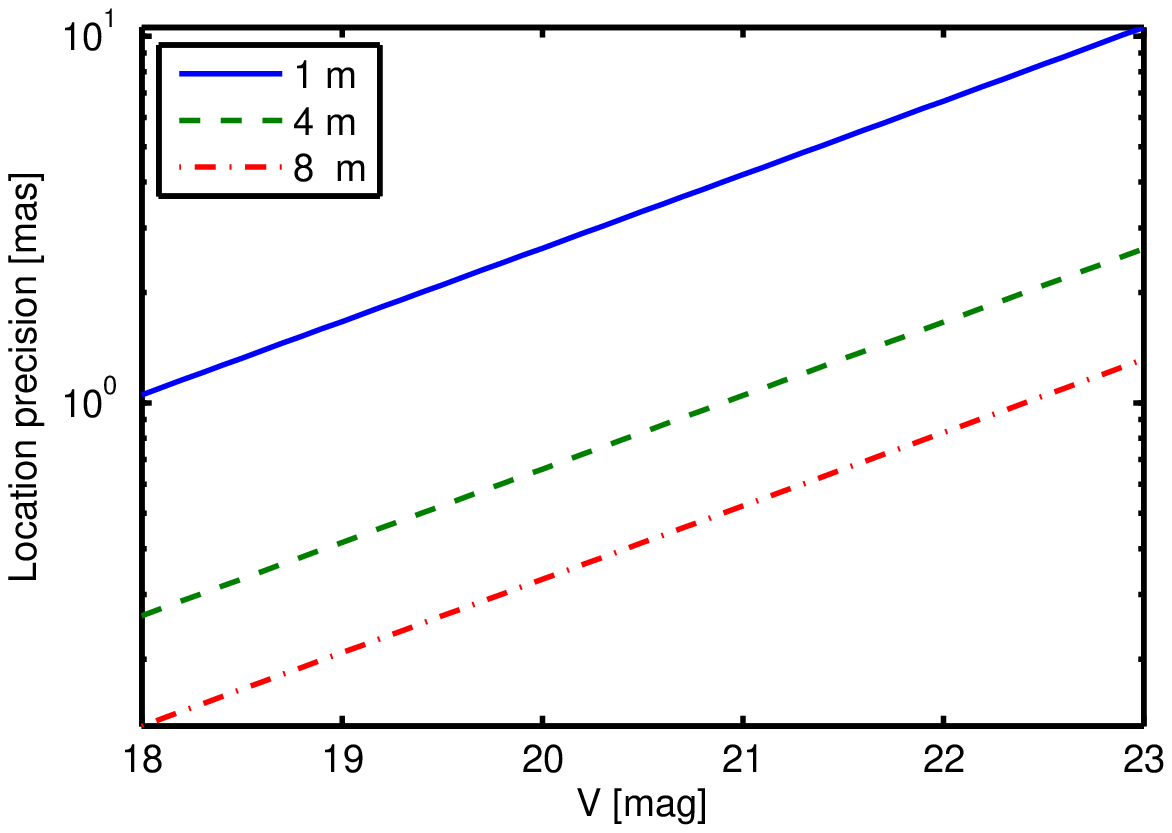}
\par\end{centering}

\caption{\label{fig:SNR_diam}SNR in one hour for different telescopes. }
\end{figure}

The following assessment is somewhat simplified in many respects,
but it is considered adequate with respect to the order of magnitude
estimation. In the elementary exposures, the image of unresolved objects
has size related to the seeing, e.g. $w\simeq1"$. The signal to noise
ratio (SNR) achieved in broadband imaging in V band, over one hour,
with different telescopes, is shown in Fig.~\ref{fig:SNR_diam} (left).
The location precision $\sigma_{p}$ as a function of the magnitude
can roughly be derived as 
\[
\sigma_{p}\simeq\frac{w}{SNR}\,,
\]
and it is shown in Fig.~\ref{fig:SNR_diam} (right). 

The order of magnitude of observing time requirements can be derived
by scaling the precision on the secondary component of UX25 to the
$15\,\mu as$ requirement from Sec.~\ref{sub:AstromSigAmpli}. We
assume an average of 8 hours per each night of observation. Even assuming
usage of a $8\, m$ telescope, the number of nights required for $3\sigma$
detection is 960, i.e. clearly not feasible within reasonable implementation
constraints.

\section{Assessment of orbit determination from space }

Several telescopes are currently being used for astrophysical observations
from space, and more advanced ones will be launched in the coming
years. Most of them are specialised in terms of wavelength, operating
mode and so on with respect to their science goal, but several offer
the option for conventional observation proposals or targets of opportunity. 

The Hubble Space Telescope (HST) has comparably small size, i.e. $2.4\, m$
diameter, but operating outside the Earth's atmosphere it can routinely
achieve near-diffraction limited performance. The preliminary assessment
refers to the UVIS channel of the WFC3 instrument, at wavelength $\sim550\, nm$,
with characteristic image width $\lambda/D=48\, mas$. This results
in a location precision on the UX25 secondary of about $0.5\, mas$
in ten minutes. A total of about $200\, hours$ seems to be required
to fulfill the science case. Even spread over several months, the
amount seems to be fairly large with respect to the existing competition. 

Gaia will provide extremely valuable constraints on our problem, in
the sense of accurate determination of the reference star frame against
which the position and motion of our target is tracked. However, due
to its fixed scan law, the sampling and exposure time are such that
the precision on the UX25 secondary motion is simply insufficient
with respect to the goal. 

The James Webb Space Telescope (JWST), planned for launch in 2018,
will provide a much better chance for implementation of the current
experiment. Given the larger size ($6.5\, m$ diameter), and higher
sensitivity, the location performance may be quite adequate: the characteristic
image width at wavelength $\lambda=650\, nm$ is $21\, mas$, resulting
in a location precision on the UX25 secondary of about $0.2\, mas$
in ten minutes. Therefore, the basic measurement sequence for orbit
determination with the required precision will require about $27\, hours$,
interspersed over the 8.3~days period. This is much more manageable
with respect to time allocation competition with other science proposals.

\section{Assessment of orbit determination from adaptive optics telescopes}

In case of AO telescopes, the available photon flux and SNR are comparable,
but the astrometric precision improves by about one order of magnitude
thanks to the higher image resolution, close to the diffraction limit.
The diffraction limit is order of $\lambda/D$, where $\lambda=1.1;\,1.65;\,2.2\,\mu m$
is the wavelength respectively in the J, H and K near infrared bands.
As a reference case, we consider the case of a $D=8\, m$ diameter
AO telescope. Then, the characteristic image size is $\sim28,\,44,\,57\, mas$,
respectively. 

\begin{figure}
\begin{centering}
\includegraphics[scale=0.55]{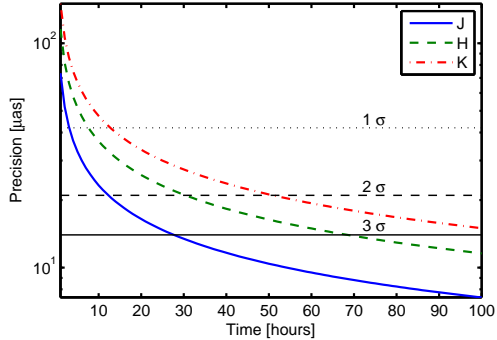}
\par\end{centering}

\caption{\label{fig:DetPerf_t_AOT8m}Detection performance vs. time}
\end{figure}

The location performance in J, H and K band vs. total exposure time
is shown in Fig.~\ref{fig:DetPerf_t_AOT8m}, respectively as solid,
dashed and dash-dot curves. The horizontal lines corresponding to
detection level $1\sigma$, $2\sigma$ and $3\sigma$ are also shown,
respectively as dotted, dashed and solid thin lines. It may be noted
that semi-ideal response is assumed, whereas additional degradation
factors are expected; e.g., the AO performance is expected to degrade
at shorter wavelength, reducing the gap between the performance curves. 

Applying the same simple SNR extrapolation as above, the number of
hours and full nights required on a semi-ideal AO telescope is listed
in Table~\ref{tab:ObsReq8mAOT}. As a baseline, K band operation
is considered. 

Even assuming conservatively that the longest wavelength were imposed
e.g. by operation constraints, it appears that quantum vacuum effects
detection at $3\sigma$ could be attained using less than two weeks
of observation. 

\begin{table}
\begin{centering}
\begin{tabular}{|c|c|c|c|c|}
\hline 
\textbf{Band} & \textbf{Wavelength {[}$\mu m${]}} & \textbf{$\lambda/D$ {[}$mas${]}} & \textbf{Time {[}hours{]}} & \textbf{Time {[}nights{]}}\tabularnewline
\hline 
\hline 
\textbf{J} & 1.1 & 28 & 24 & 3\tabularnewline
\hline 
\textbf{H} & 1.65 & 44 & 59 & 7.5\tabularnewline
\hline 
\textbf{K} & 2.2 & 57 & 100 & 12.5\tabularnewline
\hline 
\end{tabular}
\par\end{centering}

\caption{\label{tab:ObsReq8mAOT}Observation requirements on a 8 m AO telescope}
\end{table}

This is not a negligible amount of time on an advanced AO telescope,
but quite compatible with several current observing programs, so that
it may represent the reference for a realistic observing program. 

Moreover, it may be noted that the science case does not require continuous
observation, but just sparse coverage of the binary orbital period
at known times over a time elapse of order of a few months. Therefore,
service mode observations, using fractions of nights not efficiently
used by other high priority science programs, may be adopted. 

\textbf{Remark}: an AO system based on laser guide star is required,
since the target does not happen to be close enough to bright natural
stars with sufficient frequency to ensure that the required exposure
time is accumulated in a reasonable calendar time. The issue may however
be further investigated. 

\textbf{Remark}: the exposure time requirement for e.g. a $4\, m$
class AO telescope grows by a factor 4, due to both resolution and
photon budget, so that it remains fairly acceptable in J band, but
quite demanding in K band.

\section{Preliminary simulation of detection performance}

The simulation is based on least square estimation of the rotation
angle (precession) of the orbit between the beginning and end of the
observation period (5 years, corresponding to about $50\, arcsec$
precession. A number of experimental conditions have to be better
defined for a more detailed estimate. Two cases have been considered: 

- an AO equipped telescope, allowing better determination of the individual
component motion; 

- a conventional telescope, able only to ``see'' the photo-centre
motion. 

The main difference considered between AO equipped and conventional
imaging telescopes (respectively AOT and CIT) is the assumption that
the images taken in the former case, over most of the orbital period,
are resolved, thus allowing direct observation of the full angular
excursion associated to the orbital motion, whereas in the latter
case the observed quantity, i.e. the photocentre motion, only spans
a range approximately 10 times smaller. 

Given the astrometric noise associated to an elementary period of
observation, order of one night, we generate a set of noisy observations
over a period of several orbits, and evaluate the mean and standard
deviation of the estimated precession angle. 

\begin{figure}
\begin{centering}
\includegraphics[scale=0.49]{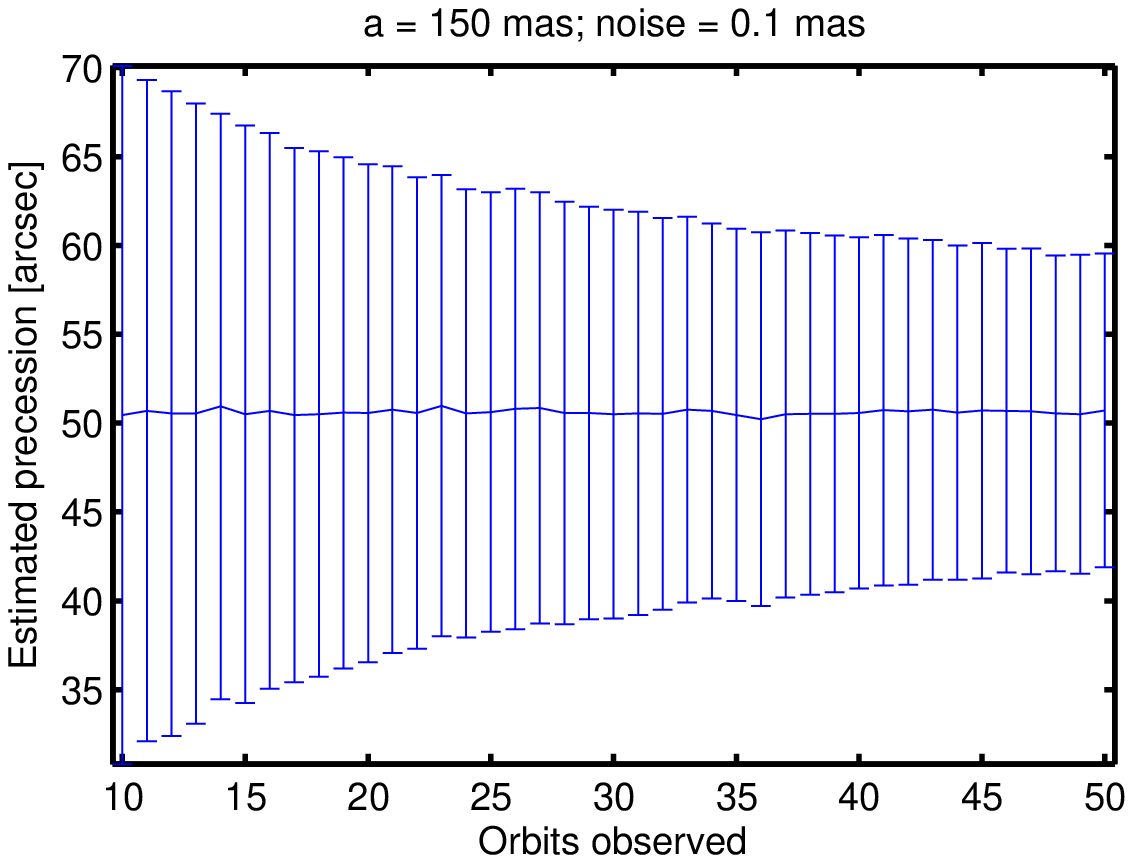}
\includegraphics[scale=0.49]{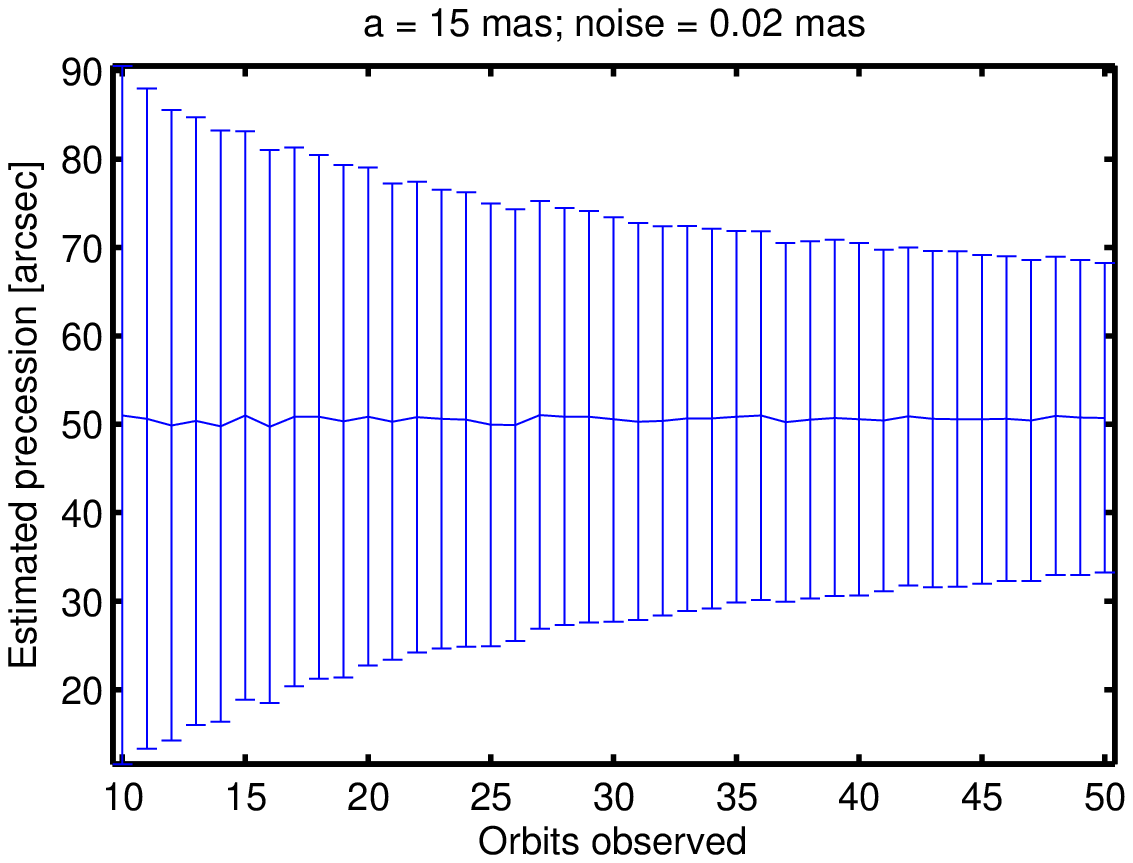}
\par\end{centering}

\caption{\label{fig:DetPerf}Detection performance of an AO telescope (left)
and a conventional telescope (right)}
\end{figure}

A meaningful detection is considered successfully achieved when, in
a few months, i.e. few ten orbits, the standard deviation of the estimated
precession is 2-3 times the average value. The assessment is qualitatively
consistent with the requirement in Sec.~\ref{sub:AstromSigAmpli}. 

As shown in Fig.~\ref{fig:DetPerf}, both an AO-equipped and a conventional
telescope are able, in principle, to detect the quantum vacuum precession
at the amplitude level mentioned in Hajdukovic (2013)\cite{hajdukovic:hal-00908554}.
However, in the latter case, an extremely low noise is required, which
could be achievable in principle by merging the observations of many
conventional telescopes over a long time.

\section{Discussion}

The preliminary assessment of astronomical quantum vacuum detection
appears to provide encouraging results, in the sense that the best
currently available instrumentation seems to be able, over a time
frame of a few years (due to the need of cumulating the precession
effects), to reach unambiguous results. 

\textbf{The observing time requirement in K band is in the range of
two weeks, for an 8 m telescope equipped with adaptive optics using
a laser guide stars (e.g. Keck), possibly spread over several months. }

Usage of conventional ground based telescopes appears to be quite
impractical. 

The HST may achieve the proposed goal, but the total exposure time
required seems to raise doubts on practical feasibility. 

The issue requires further investigation in terms of observation improvement
based on the considerations from Sec.~\ref{sub:ExpApproach}, and
the characterization of the UX25 binary system should be addressed
in more detail.

\bibliographystyle{plain}
\bibliography{note}

\end{document}